\documentstyle[12pt]{article}
\def\beq{\begin{equation}}
\def\eeq{\end{equation}}
\def\bea{\begin{eqnarray}}
\def\eea{\end{eqnarray}}
\def\bq{\begin{quote}}
\def\eq{\end{quote}}

\def\gappeq{\mathrel{\rlap {\raise.5ex\hbox{$>$}}
{\lower.5ex\hbox{$\sim$}}}}

\def\lappeq{\mathrel{\rlap{\raise.5ex\hbox{$<$}}
{\lower.5ex\hbox{$\sim$}}}}

\begin{document}
\topmargin -0.5cm
\oddsidemargin -0.3cm
\pagestyle{empty}
\begin{flushright}
{HIP-1999-41/TH}
\end{flushright}
\vspace*{15mm}
\begin{center}
{\bf Third-order phase transition and superconductivity  in thin films} \\
\vspace*{1.5cm} 
{\bf Christofer Cronstr\"{o}m}$^{*)}$ \\
\vspace{0.3cm}
Division of Theoretical Physics, Physics Department\\
P. O. Box 9, FIN-00014 University of Helsinki; Finland \\
\vspace{0.5cm}
and\\
\vspace{0.5cm}
{\bf Milan Noga}$^{\dagger)}$  \\
\vspace{0.3cm}
Department of Theoretical Physics, Comenius University\\
Mlynska Dolina, 84215 Bratislava, Slovak Republic \\
\vspace*{1cm}  
{\bf ABSTRACT} \\ \end{center}
\vspace*{5mm}
\noindent
We have found a new mean field solution in the BCS theory of superconductivity. This
unconventional solution indicates the existence of superconducting phase transitions of third
order in thin films, or in bulk matter with a layered structure. The critical temperature
increases with decreasing  thickness of the layer, and does not exhibit the isotope effect. The
electronic specific heat is a continuous function of temperature with a discontinuity in its 
derivative. 

\vspace*{1cm} 
\noindent 
%\rule[.1in]{16.5cm}{.002in}

\noindent
$^{*)}$ e-mail address: Christofer.Cronstrom@Helsinki.fi\\
$^{\dagger)}$ e-mail address: Milan.Noga@fmph.uniba.sk\\
%\vspace*{2cm}
%\begin{flushleft}HU-TFT XXXX/96 \\
%January 1996
%\end{flushleft}
\vfill\eject

%\pagestyle{empty}
%\clearpage\mbox{}\clearpage

\setcounter{page}{1}
\pagestyle{plain}

\section{Introduction}  % Produces section heading.  Lower-level
                                    % sections are begun with similar 
                                    % \subsection and \subsubsection commands.

The BCS theory of superconductivity \cite{BCS} has been very succesful in explaining 
properties of a large class of simple superconductors in terms of just two experimental
parameters, namely the squared electron-phonon coupling constant $g$ and the Debye 
frequency $\omega_{D}$. The BCS theory in its simplest form is based on the following so-called
BCS reduced grand canonical Hamiltonian,
\begin{equation}
H = \sum_{\bf{k},\sigma} \xi_{\bf{k}}a^{\dag}_{\bf{k} \sigma}a_{\bf{k} \sigma}
-\frac{g}{V} \sum_{\bf{k}, \bf{k'}}\,\Theta(\hbar\omega_{D} - \mid \xi_{{\bf k}}\mid ) \Theta(\hbar\omega_{D} - \mid \xi_{{\bf k'}}\mid ) a^{\dagger}_{\bf{k},+}a^{\dagger}_{\bf{-k},-}
a_{\bf{-k'},-}a_{\bf{k'},+} 
\label{eq:1}
\end{equation}
%\ref{eq:1}
where the quantities $\bf{k}$ and $\bf{k'}$ are wave vector variables, the quantity $\sigma = \pm$
is the spin projection, and the symbols $a_{\bf{k} \sigma}$ and $a^{\dagger}_{\bf{k} \sigma}$
are electron field operators. Furthermore,
\begin{equation}
\xi_{\bf{k}} = \frac{\hbar^{2}\bf{k}^{2}}{2m} - \mu \label{eq:1a}
\end{equation} 
where $\mu $ is the chemical potential and $m$ is the electron mass. Finally, $V$ is the volume
of the system. The sum over $\bf{k}$ and $\bf{k'}$ in Eq. (\ref{eq:1}) is restricted by the 
conditions $ |\xi_{\bf{k}}| \leq \hbar \omega_{D}$ and $|\xi_{\bf{k'}}| \leq \hbar \omega_{D}$,
as indicated by the appropriate stepfunctions $\Theta$ in Eq. (\ref {eq:1}).

The Hamiltonian (\ref{eq:1}) represents by itself a great simplification of the net interaction
between the electrons. Even so, in BCS theory one makes an additional technical simplification
by adopting merely the mean field approximation in calculating physical properties of
superconductors \cite{Abrikosov}. These simplifications notwithstanding, the BCS theory predicts
a number of relations, some of which are even independent of the phenomenological parameters
$g$ and $\omega_{D}$, which are surprisingly well obeyed by a large class of superconductors
\cite{Meservey}.

As is well known, before the BCS theory was developed, Ginzburg and Landau had proposed a theory
of superconductivity, currently known as the Ginzburg-Landau theory \cite{Abrikosov 2}. The 
equations of this theory were derived from the phenomenological Landau theory of second-order 
phase transitions \cite{Landau}. In accordance with the Landau theory, in all cases of second
order phase transitions, one has to define the so-called order parameters, which characterise
ordered structures of macroscopic fields appearing spontaneously in systems below a critical
temperature $T_{c}$. Thus the order parameters equal zero for $T > T_{c}$ and acquire non-vanishing
values for $T < T_{c}$. Later Gor'kov demonstrated \cite{Gor'kov} by using the Green function
method, that the Ginzburg-Landau theory is a limting case of the BCS theory, provided that the
order parameters are values of a complex function $\varphi(\bf{x})$, which is proportional to 
the anomalous Green function $<\psi_{-}(\bf{x})\psi_{+}(\bf{x})>$ of the electron field operators
$\psi_{\sigma}(\bf{x})$, evaluated in the mean field approximation. The expression for the 
function $\varphi(\bf{x})$ is thus the following,
\begin{equation}
\varphi({\bf{x}}) = C<\psi_{-}({\bf{x}})\psi_{+}({\bf{x}})>_{mf} = \frac{C}{V}
\sum_{{\bf{k}}, {\bf{k'}}}\!^{\prime}\, <a_{\bf{k},-}a_{\bf{k'},+}>_{mf} \exp i(\bf{k}+\bf{k'})
\cdot \bf{x}\, , \label{eq:2}
\end{equation}
where $<...>_{mf}$ denotes the statistical grand canonical ensemble average with the Hamiltonian
(\ref{eq:1}) in the mean field approximation. From Eq. (\ref{eq:2}) one sees that ordered 
superconducting structures are characterised by the mean field correlation functions 
$<a_{\bf{k},-}a_{\bf{k'},+}>_{mf}$ corresponding to the Hamiltonian (\ref{eq:1}).
Thus, the order parameters will be chosen to be related to these mean field correlation functions.

A very important theorem concerning mean field solutions to the BCS theory of superconductivity has 
been proved by Bogoliubov \cite {Bogolu}. This theorem states that any mean field solution of the 
BCS Hamiltonian (\ref {eq:1}) becomes an exact solution in the thermodynamic limit. This means
that the effects of quantal fluctuations about a given mean field give no contributions to 
thermodynamical potentials. This was also proven explicitly in \cite {Milan} by a direct evaluation 
of the grandcanonical partition function corresponding to the Hamiltonian (\ref {eq:1}).

Although the BCS theory has existed for almost fourty years, apparently only one of its mean 
field solutions corresponding to the Hamiltonian (\ref{eq:1}) has been considered so far. To the 
best of our knowledge, there have been no attempts to explore the possibility of finding more 
than one mean field solution with the Hamiltonian (\ref{eq:1}).

In this paper we report on our observation that there is at least one additional mean field solution
to the system described by the Hamiltonian (\ref{eq:1}). We describe this solution explicitly
below. The superconducting state corresponding to the new mean field solution can appear at non-zero
temperature only in quasi-twodimensional systems such as thin films or layered structures of 
bulk material. Such layered structures are actually present in high $T_{c}$ superconductors
\cite{Burns}. The new mean field solution described in this paper corresponds to a phase 
transition to a superconducting state with some rather unexpected properties. Specifically,
it is a third-order phase transition which does not exhibit the isotope effect which is typical
for the usual BCS theory. Thus, this phase transition cannot be described by the ordinary 
phenomenological Ginzburg-Landau theory, which describes phase transitions of  second order.

\section{The mean field approximation and the gap equation}
%\vskip 1 cm
In order to analyse the novel mean field solution to the BCS theory with the Hamiltonian (\ref {eq:1})
we first express the interaction Hamiltonian $H_{I}$ in (\ref {eq:1}) as a sum of two terms,
\begin{equation}
H_{I} = H'_{I} + H''_{I}
\label{eq:twoterm}
\end{equation}
where
\begin{equation}
H'_{I} = -\frac{g}{V} \sum_{\bf{k}, \bf{k'}}\,\Theta(\hbar\omega_{D} - \mid \xi_{{\bf k}}\mid ) \Theta(\hbar\omega_{D} - \mid \xi_{{\bf k'}}\mid ) \delta_{\mid {\bf k}\mid, \mid {\bf k'}\mid}\,a^{\dagger}_{\bf{k},+}a^{\dagger}_{\bf{-k},-}
a_{\bf{-k'},-}a_{\bf{k'},+}
\label{eq:four}
\end{equation}
and
\begin{equation}
H''_{I} = -\frac{g}{V} \sum_{\bf{k}, \bf{k'}} \,\Theta(\hbar\omega_{D} - \mid \xi_{{\bf k}}\mid ) \Theta(\hbar\omega_{D} - \mid \xi_{{\bf k'}}\mid ) (1 - \delta_{\mid {\bf k}\mid, \mid {\bf k'}\mid})a^{\dagger}_{\bf{k},+}a^{\dagger}_{\bf{-k},-}
a_{\bf{-k'},-}a_{\bf{k'},+}
\label{eq:five}
\end{equation}

%The sum over ${\bf k}$ and ${\bf k'}$ in Eq. (\ref {eq:five}) is restricted by the condition
%$\mid {\bf k'} \mid \neq \mid {\bf k} \mid $ as indicated by the prime in the summation sign
%($\sum\!^{\prime}$).

In the interaction term defined by Eq. (\ref {eq:four}) we introduce new summation indices
${\bf p}$ and ${\bf q}$ related to ${\bf k}$ and ${\bf k'}$ as follows,

\begin{equation}
{\bf p} = \frac{{\bf k}Ê+ {\bf k'}}{2}\,,\; {\bf q} = \frac{{\bf k}Ê- {\bf k'}}{2} 
\label{eq:six}
\end{equation}
The innocent looking change of summation variables introduced in (\ref {eq:six}) is very important,
as it enables one to satisfy the condition $\mid {\bf k} \mid = \mid {\bf k'} \mid$ in the sum
in Eq. (\ref {eq:four}) by any two perpendicular vectors ${\bf p}$ and ${\bf q}$. This will be seen
to be quite essential for the definition of order parameters below.

The Hamiltonian (\ref {eq:1}) can now be written as follows,  
\begin{equation}
H = \sum_{\bf{k},\sigma} \xi_{\bf{k}}a^{\dag}_{\bf{k} \sigma}a_{\bf{k} \sigma}
-\frac{g}{V} \sum_{\bf{p}, \bf{q}}\,\Theta(\hbar\omega_{D} - \mid \xi_{{\bf p} + {\bf q}}\mid ) \delta_{\mid {\bf p} + {\bf q} \mid, \mid {\bf p} - {\bf q} \mid}\,a^{\dagger}_{{\bf p} + {\bf q},+}a^{\dagger}_{{-\bf p}-{\bf q},-}
a_{{-\bf p} + {\bf q},-}a_{{\bf p} - {\bf q},+} + H''_{I}
\label{eq:seven}
\end{equation}
Let us note that it is enough to retain only one step function in the equation (\ref {eq:seven})
above, since $\xi_{{\bf p} + {\bf q}} = \xi_{{\bf p} - {\bf q}}$ in the sum in Eq. 
(\ref {eq:seven}).
 
We next introduce a general set of complex order parameters $b_{\bf{q,p}}$ and $b^{\ast}_{\bf{q,p}}$
by the following definitions, 
\begin{equation}
b_{\bf{q,p}} = g <a_{\bf{-(p+q)},-}a_{\bf{(p+q)},+}>\delta_{\mid{\bf p} + {\bf q} \mid, \mid {\bf p} - {\bf q} \mid} \;,\;
b^{\ast}_{\bf{q,p}}  = g <a^{\dagger}_{\bf{(p+q)},+}a^{\dagger}_{\bf{-(p+q)},-}>\delta_{\mid{\bf p} + {\bf q} \mid, \mid {\bf p} - {\bf q} \mid} 
\label{eq:eight}
\end{equation}
\noindent
The order parameters $b_{\bf{q,p}}$ and $b^{\ast}_{\bf{q,p}}$ are thus enumerated by two orthogonal vectors
${\bf p}$ and ${\bf q}$, which are restricted by the condition 
$\mid \xi_{{\bf p} + {\bf q}}\mid = \mid \xi_{{\bf p} - {\bf q}}\mid \leq \hbar\omega_{D}$.

We now define the following gap functions $\Delta_{\bf{k}}$ and $\Delta^{\ast}_{\bf{k}}$ in 
terms of the functions $b_{\bf{q,p}}$ and $b^{\ast}_{\bf{q,p}}$ introduced above,
\begin{equation}
\Delta_{\bf{k}} = \frac{1}{V} \sum_{\bf{p,q}} b_{\bf{q,p}} \delta_{\bf{p-q,k}},\;\;    
\Delta^{\ast}_{\bf{k}} = \frac{1}{V} \sum_{\bf{p,q}} b^{\ast}_{\bf{q,p}}
\delta_{\bf{p-q,k}} 
\label{eq:nine}
\end{equation}
\noindent
Using the definitions (\ref{eq:eight}) and (\ref{eq:nine}), we decompose the Hamiltonian 
(\ref{eq:seven}) as a sum of two terms,
\begin{equation}
H = H_{mf} + H_{fl}, 
\label{eq:5a}
\end{equation}
where
\begin{eqnarray}
\label{eq:ten}
H_{mf} & = & \frac{1}{gV} \sum_{\bf{p,q}}b^{\ast}_{\bf{q,p}}b_{\bf{-q,p}} + 
\nonumber\\
       &   &+  \sum_{\bf{k}}\left \{ \xi_{\bf{k}}(a^{\dag}_{\bf{k},+}a_{\bf{k},+} +
a^{\dag}_{\bf{k},-}a_{\bf{k},-}) - (\Delta^{\ast}_{\bf{k}}a_{\bf{-k},-}a_{\bf{k},+} +
\Delta_{\bf{k}}a^{\dag}_{\bf{k},+}a^{\dag}_{\bf{-k},-}) \right \} \label{eq:6}
\end{eqnarray}
is the mean field Hamiltonian, and the term
\begin{eqnarray}
\label{eq:eleven}
H_{fl} & = & -\frac{g}{V} \sum_{\bf{p}, \bf{q}}\!^{\prime}\,\Theta(\hbar\omega_{D} - \xi_{{\bf p} + {\bf q}})\left [a^{\dagger}_{\bf{(p+q)},+}
a^{\dagger}_{\bf{-(p+q)},-} - <a^{\dagger}_{\bf{(p+q)},+}a^{\dagger}_{\bf{-(p+q)},-}>
\right ]\nonumber\\
       &   &\left [a_{\bf{-(p-q)},-}a_{\bf{(p-q)},+} - <a_{\bf{-(p-q)},-}a_{\bf{(p-q)},+}> 
\right ]\delta_{\mid {\bf p} + {\bf q} \mid, \mid {\bf p} - \bf q} \mid  + H''_{I}
\end{eqnarray}
describes quantal fluctuations about the mean fields $b_{\bf{q,p}}$ and $b^{\ast}_{\bf{q,p}}$, 
cf. Eq. (\ref{eq:eight}).

In view of the Bogoliubov theorem quoted above, the quantal fluctuation Hamiltonian 
(\ref {eq:eleven}) has no macroscopic effects in the thermodynamical limit. By neglecting
 it in all expressions, i.e.  replacing
the Hamiltonian (\ref{eq:seven}) with the mean field Hamiltonian (\ref{eq:ten}) one gets all physical
quantities in the so-called mean-field approximation. From now on we use this approximation
throughout.

The Hamiltonian $H_{mf}$, eq. (\ref{eq:ten}), can easily be diagonalized, and results then in the
following partition function $Z_{mf}$,
\begin{eqnarray}
Z_{mf} & \equiv & Tr \exp \left [- \beta H_{mf} \right ] \nonumber\\
       & = & \exp \left \{ - \frac{\beta}{gV} \sum_{\bf {p,q}}b^{\ast}_{\bf{q,p}}
b_{\bf{-q,p}}\right \}\left \{ \prod_{\bf {k}} \left (e^{-\beta \xi_{\bf {k}}}
(1 + e^{\beta E_{\bf {k}}})(1 + e^{-\beta E_{\bf {k}}})\right ) \right \}, \label{eq:7}
\end{eqnarray}
where
\begin{equation}
E_{\bf {k}} = \left ( \xi^{2}_{\bf {k}} + |\Delta_{\bf {k}}|^{2} 
\right )^{\frac{1}{2}}
\label{eq:8}
\end{equation}
is the quasiparticle energy spectrum.

Evaluating the correlation functions in the mean field approximation, one then gets the following
relations from the equations (\ref{eq:eight}),
\begin{equation}
b_{\bf{q,p}} = \frac{g}{2E_{\bf {p+q}}} \Delta_{\bf {p+q}}\; th \left ( \frac{1}{2} \beta E_{\bf {p+q}} \right ). 
\label{eq:9}
\end{equation}
\noindent
The sets of equations (\ref{eq:9}) for the determination of the functions $b_{\bf{q,p}}$ and
$b^{\ast}_{\bf{q,p}}$ are closedly related to the so-called gap equations in the standard
BCS-theory of supeconductivity, as will be seen presently.

\section{The solutions $b_{\bf{q,p}}$ and $b^{\ast}_{\bf{q,p}}$}
%\vskip 1 cm
Using now equations (\ref{eq:nine}) and (\ref{eq:9}), one gets the following equation for the 
gap function $\Delta_{\bf {k}}$,
\begin{equation}
\Delta_{\bf {k}} = \frac{g}{V}\sum_{\bf{p,q}}\!^{\prime}\,
\frac{\Delta_{\bf {p+q}}}{2E_{\bf {p+q}}}\; th \left (\frac{1}{2} \beta E_{\bf {p+q}}\right )
\delta_{\bf {p-q, k}}\delta_{|\bf {p-q}|,|\bf {p+q}|}\Theta(\hbar\omega_{D} - \mid \xi_{{\bf p} + {\bf q}} \mid)
\label{eq:15}
\end{equation}
We now revert to the summation indices ${\bf k'} = {\bf p} + {\bf q}$, ${\bf k''} = {\bf p} - {\bf q}$
and carry out the summation over ${\bf k''}$ with the result
\begin{equation}
\Delta_{{\bf k}} = \frac{g}{2V} \sum_{{\bf k'}} \frac{\Delta_{{\bf k'}}}{E_{{\bf k'}}}\; th \frac{\beta}{2} E_{{\bf k'}}\; \delta_{\mid {\bf k'}\mid, \mid {\bf k}\mid}\, \Theta (\hbar\omega_{D} - \mid \xi_{{\bf k'}}\mid)
\label{eq:sixteen}
\end{equation}
The assumption that the gap function $\Delta_{{\bf k'}}$ is a function of the magnitude $\mid {\bf k'} \mid$
only, simplifies the relation (\ref{eq:sixteen}) substantially, as follows,
\begin{equation}
\Delta_{{\bf k}} = \frac{g N_{1}}{2V}\frac{\Delta_{{\bf k}}}{E_{{\bf k}}}\, th \frac{\beta}{2} E_{{\bf k}} 
\label{eq:seventeen}
\end{equation}
where
\begin{equation}
N_{1} \equiv \sum_{{\bf k'}} \delta_{\mid {\bf k'} \mid, \mid {\bf k} \mid } \Theta(\hbar\omega_{D} - \mid \xi_{{\bf k'}}\mid )
\label{eq:eighteen}
\end{equation}
\noindent
The quantity $N_{1}$ defined in Eq. (\ref {eq:eighteen}) is the number of wave vectors ${\bf k'}$
with given length $\mid {\bf k'} \mid = \mid {\bf k } \mid $. The vectors ${\bf k'}$ have
the components given below,
\begin{equation}
{\bf k'} = \left ( \frac{2\pi}{L_{1}} n_{1} , \frac{2 \pi}{L_{2}} n_{2} , \frac{2 \pi}{L_{3}} n_{3} \right )
\label{eq:nineteen}
\end{equation}
where $n_{1}$, $n_{2}$ and $n_{3}$ are integers, $L_{1}$, $L_{2}$ and  $L_{3}$ are the lengths
of the edges of the box into which the system is enclosed. The condition 
$\mid {\bf k'} \mid =  \mid {\bf k} \mid$ defines the surface of an ellipsoid with semiaxes 
$a_{1} = \frac{kL_{1}}{2\pi}$, $a_{2} = \frac{kL_{2}}{2\pi}$, and $a_{3} = \frac{kL_{3}}{2\pi}$,
i.e.
\begin{equation} 
\frac{n_{1}^{2}}{a_{1}^{2}} + \frac{n_{2}^{2}}{a_{2}^{2}} + \frac{n_{3}^{2}}{a_{3}^{2}} = 1
\label{eq:twenty}
\end{equation}
\noindent 
Thus the number of states $N_{1}$ is in fact equal to the surface area of the ellipsoid 
(\ref {eq:twenty}) which is determined by an elliptic integral. However for a very thin film, 
or layer with $L_{1} \ll L_{2}$, $L_{1} \ll L_{3}$ the number of states can be approximated by
the surface area of a very thin disc of elliptical shape, with semiaxes $a_{2} = \frac{kL_{2}}{2\pi}$
and $a_{3} = \frac{kL_{3}}{2\pi}$. Thus the number of states $N_{1}$ is given by the following
expression,
\begin{equation}
N_{1} = 2\pi k^{2} \frac{L_{2}L_{3}}{(2\pi)^{2}} = \frac{1}{2\pi} k^{2} L_{2}L_{3}
\label{eq:twentyone}
\end{equation}
Since the vectors ${\bf k'}$ in the sum (\ref {eq:sixteen}) are restricted by the inequalities 
below,
\begin{equation}
{\cal E}_{F} - \hbar\omega_{D} \leq \frac{\hbar{\bf k'}^{2}}{2m} \leq {\cal E}_{F} + \hbar\omega_{D} 
\label{eq:ineq21.5}
\end{equation}
and since $\hbar\omega_{D} \ll {\cal E}_{F}$ we can replace the quantity $k^{2}$ in 
(\ref {eq:twentyone}) by $k_{F}^{2}$, i.e.
\begin{equation}
N_{1} = \frac{1}{2\pi} k_{F}^{2} L_{2} L_{3}
\label{eq:twentytwo}
\end{equation}
By using the expression for the density of states of electrons with a single spin projection
at the Fermi level $N_{0}$,
\begin{equation}
N_{0} = \frac{2mk_{F}}{(2\pi\hbar)^{2}}
\label{eq:ennzero}
\end{equation}
we can express the ratio $\frac{N_{1}}{2V}$ as follows,
\begin{equation}
\frac{N_{1}}{2V} = N_{0} \sqrt{{\cal E}_{F}} \frac{\hbar}{\sqrt{2m}} \frac{\pi}{L_{1}}
\label{eq:twentythree}
\end{equation}

Using the last relation in Eq. (\ref {eq:seventeen}) we obtain the formula,
\begin{equation}
\Delta_{\bf {k}} = \Delta_{\bf {k}} g N_{0} \sqrt{{\cal E}_{F}{\cal E}_{1}} \frac{1}{E_{\bf k}}
\; th \frac{1}{2} \beta E_{\bf k}\;, 
\label{eq:17}
\end{equation}
where
\begin{equation}
{\cal E}_{1} = \frac{\hbar^{2}}{2m}\;\frac{\pi^{2}}{L^{2}_{1}} 
\label{eq:17a}
\end{equation}
is the lower bound of the energy spectrum of free electrons in a thin box of volume 
$V = L_{1}L_{2}L_{3}$  with $L_{2} \gg L_{1}$ and $L_{3} \gg L_{1}$. From eq. (\ref{eq:17})
follows that the energy spectrum $E_{\bf k}$ of the quasiparticles in the superconducting state 
with $\Delta_{\bf {k}}$ $\not= 0$ must satisfy the following relation,
\begin{equation}
1 =  g N_{0} \sqrt{{\cal E}_{F}{\cal E}_{1}} \frac{1}{E_{\bf k}}
\; th \frac{1}{2} \beta E_{\bf k}\;, \label{eq:18}
\end{equation}
which equation can be satisfied only if the temperature $T$ is bounded from above by the 
following critical temperature $T_{c}$,
\begin{equation}
T_{c} = \frac{1}{2k_{B}} g N_{0}\sqrt{{\cal E}_{F}{\cal E}_{1}} \;. \label{eq:19}
\end{equation}

The critical temperature $T_{c}$ defined by eq. (\ref{eq:19}) does not depend on the Debye 
frequency $\omega_{D}$ and hence it does not exhibit the isotope effect typical for the
standard solution in BCS theory. The critical temperature also depends on the lowest bound ${\cal E}_{1}$ of the energy spectrum 
of the electrons in the superconducting material. For bulk materials ${\cal E}_{1} \rightarrow 0$
so this kind of superconductivity does not exist in bulk samples. However for thin films the lower
bound ${\cal E}_{1}$ is nonzero and given by Eq. (\ref {eq:17a}). The critical temperature increases 
with decreasing thickness of the layer.

The solution to eq. (\ref{eq:18}) for the energy spectrum $E_{\bf k}$ in the range of ${\bf k}$
for which $\Delta_{\bf k}$ $\not= 0$, is a quantity $\eta(T)$ say, which is dependent on the 
temperature $T$ only,
\begin{equation}
E_{\bf k} = (\xi^{2}_{\bf k} + |\Delta_{\bf k}|^{2})^{\frac{1}{2}} \equiv \eta(T) \label{eq:20}
\end{equation}
One cannot obtain an explicit expression for the solution $\eta(T) = E_{\bf k}$ of eq. 
(\ref{eq:18}) for general values of $T$ in the range $T < T_{c}$ but has to resort to 
numerical methods for these general values of $T$. However in the limiting cases
$T \ll T_{c}$ and $T_{c} - T \ll T_{c}$, respectively, one readily obtains the following
formulae from eq. (\ref{eq:18}),
\begin{equation}
\eta(T) \approx \eta_{0} (1 - 2 e^{-\beta \eta_{0}})\; ; T \ll T_{c}, \label{eq:21}
\end{equation}
and
\begin{equation}
\eta(T) \approx \sqrt{3} \eta_{0} \left (1- \frac{T}{T_{c}} \right )^{\frac{1}{2}}\;;
T_{c} - T \ll T_{c}\;, \label{eq:21a}
\end{equation}
where
\begin{equation}
\eta_{0} = \eta(T = 0) = 2k_{B}T_{c} . \label{eq:22}
\end{equation}
The ratio
\begin{equation}
\frac{\eta_{0}}{k_{B}T_{c}} = 2 \label{eq:22a}
\end{equation}
is a universal constant, independent of the physical properties of the thin layer under
consideration.

From eq. (\ref{eq:20}) we get the following expression for the gap function $\Delta_{\bf k}$,
\begin{equation}
\Delta_{\bf k} = \left ( \eta^{2} - \xi^{2}_{\bf k} \right )^{\frac{1}{2}} 
\Theta (\eta - |\xi_{\bf k}|). \label{eq:23}
\end{equation}
Combining eq. (\ref{eq:23}) with (\ref{eq:9}) one finally obtains an explicit expression for
the order parameter $b_{\bf {q,p}}$,
\begin{equation}
b_{\bf {q,p}} = b_{\bf {-q,p}} = \frac{g}{2\eta_{0}} \left \{ \eta^{2} - \left [ \frac{\hbar^{2}}
{2m} ({\bf {q}}^{2} + {\bf {p}}^{2}) - {\cal E}_{F} \right ]^{2} \right \}^{\frac{1}{2}}
\Theta(\eta - |\frac{\hbar^{2}}{2m} ({\bf {q}}^{2} + {\bf {p}}^{2}) - {\cal E}_{F}|). \label{eq:24}
\end{equation}
The last two formulae, (\ref{eq:23}) and (\ref{eq:24}), are of course consistent with the
definition (\ref{eq:nine}). The formulae (\ref{eq:18}) - (\ref{eq:24}) are valid under the following
condition,
\begin{equation}
\eta(T) \leq \hbar \omega_{D} \label{eq:25}
\end{equation}
because for $|\xi_{\bf k}| \geq \hbar \omega_{D}$ the quasiparticle energy spectrum $E_{\bf k}$
must reduce to the energy spectrum of free electrons. Combining the relation (\ref{eq:22}) with 
the inequality (\ref{eq:25}) one obtains,
\begin{equation}
T_{c} \leq \frac{\hbar \omega_{D}}{2k_{B}} \equiv \frac{1}{2}\, T_{D} 
\label{eq:26}
\end{equation}
where $T_{D}$ is the Debye temperature. For the case under consideration we thus have an upper
bound for the critical temperature. Critical temperatures $T_{c}$ in the vicinity of this upper 
bound have in fact been observed in layered structures of high $T_{c}$ superconductors 
\cite{Burns}.

The expressions (\ref{eq:23}) and (\ref{eq:24}) indicate unusual properties of the corresponding
superconducting state. Namely, not all of the order parameters $b_{\bf {q,p}}$ with 
${\bf p \cdot q} = 0$ acquire spontaneously non-vanishing values at the same time when $T < T_{c}$. For $T = 
T_{c} - 0$ only one order parameter is non-vanishing at the opening of the energy gap
$\eta(T = T_{c} - 0) = 0+$. Decreasing the temperature from the value $T_{c}$ the gap in the
energy spectrum $\eta(T)$ opens up more and more, whence more and more order parameters
acquire non-vanishing values for $T < T_{c}$. 

\section{Thermodynamical properties}
%\vskip 1 cm 
We next calculate thermodynamical properties related to the novel solution analysed above.
Using the standard method \cite{Abrikosov} we express the grand canonical potential 
$\Omega_{s}(T)$ of the superconducting state by the formula
\begin{equation}
\Omega_{s}(T) - \Omega_{n}(T) = \frac{2}{3} VN_{0} \int_{0}^{\eta} d\eta^{\prime}
(\eta^{\prime})^{3} \frac{d}{d\eta^{\prime}} \left [\frac{1}{\eta^{\prime}}\; th \frac{1}{2}
\beta \eta^{\prime} \right ] \;, \label{eq:27}
\end{equation}
where $\Omega_{n}(T)$ is the normal-state grand canonical potential of the system if it were
in a normal state at temperature $T$. The integral (\ref{eq:27}) requires numerical analysis
for general values of $T$ in the appropriate range. However, similarly as in the previous cases
(\ref{eq:21}) and (\ref{eq:21a}) the following limiting cases are readily obtained,
\begin{equation}
\Omega_{s}(T) - \Omega_{n}(T) = - \frac{1}{3} VN_{0} \eta^{2}_{0} \left [1 - \frac{\pi^{2}}{4}
\left (\frac{T}{T_{c}} \right )^{2} \right ] \;, T \ll T_{c} \label{eq:28}
\end{equation}
and
\begin{equation}
\Omega_{s}(T) - \Omega_{n}(T) = - \frac{4\sqrt{3}}{5} VN_{0} \eta^{2}_{0} 
\left (1 - \frac{T}{T_{c}} \right )^{\frac{5}{2}} \;, T_{c} - T \ll T_{c} \label{eq:29}
\end{equation}
Using the formulae (\ref{eq:28}) and (\ref{eq:29}) one can calculate the electronic specific
heat $C_{s}(T)$ and the critical magnetic field $H_{c}(T)$ with the following results,
\begin{equation}  
C_{s}(T) = 4VN_{0} \eta_{0} k_{B} (\beta \eta_{0})^{2} e^{- \beta \eta_{0}}\;, 
T \ll T_{c}\;, \label{eq:30}
\end{equation}
\begin{equation}  
H_{c}(T) = \left ( \frac{8\pi}{3}N_{0} \right )^{\frac{1}{2}} \eta_{0}
 \left [1 - \frac{\pi^{2}}{8} \left (\frac{T}{T_{c}} \right )^{2} \right ]\;, 
T \ll T_{c}. \label{eq:30a}
\end{equation}
Likewise, in the limiting case corresponding to eq. (\ref{eq:21a}) one obtains, 
\begin{equation}
\frac{C_{s}(T) - C_{n}(T)}{C_{n}(T)} = \frac{6\sqrt{3}}{\pi^{2}}
\left (1 - \frac{T}{T_{c}}\right )^{\frac{1}{2}} \;, T_{c} - T \ll T_{c} \label{eq:31} 
\end{equation}
as well as
\begin{equation}
H_{c}(T) = \left ( \frac{32\pi\sqrt{3}}{5} N_{0} \right )^{\frac{1}{2}} \eta_{0}
\left (1 - \frac{T}{T_{c}} \right )^{\frac{5}{4}} \;, T_{c} - T \ll T_{c} \label{eq:32}        
\end{equation}
where
\begin{equation}
C_{n}(T) = \frac{2\pi^{2}}{3} VN_{0} k_{B}^{2} T \label{eq:32a}
\end{equation}
is the normal-state electronic specific heat.

In the low-temperature region $T \ll T_{c}$ the formulae (\ref{eq:28}) - (\ref{eq:32}) do not
yield results which differ appreciably from the standard solution \cite {Abrikosov}. However,
near the critical temperature, i.e. for $T_{c} - T \ll T_{c}$ the differences are crucial.
The formulae (\ref{eq:29}), (\ref{eq:31}) and (\ref{eq:32}) yield critical exponents which differ
from those known for the standard solution. The electronic specific heat given by eq. 
(\ref{eq:31}) is a continuos function of $T$ at $T = T_{c}$, however with a discontinuouss
derivative at $T = T_{c}$, in contradistinction to the standard case. The novel mean field
solution considered here gives rise to a superconducting phase transition of {\em third order},
and can therefore not be described by the ordinary macroscopic Ginzburg-Landau theory of 
superconductivity, which is relevant for the phase transitions of second order.

\section{Summary}
%\vskip 1 cm 
The unconventional superconducting state which we have found within the mean field treatment
of BCS theory can appear in thin films, or in layered structures in bulk material. Therefore,
the existence of this novel feature of BCS theory can have relevance for high $T_{c}$ 
superconductors or for the explanation of the fact that even some simple metals can become 
superconducting as thin films, but not as bulk material. The fact that the BCS Hamiltonian
contains the possibility of two different superconducting structures, namely an unconventional
superconducting state at a critical temperature $T_{c1}$, and the well-known superconducting
state at a critical temperature $T_{c}$, offers the possibility of observing
consecutive superconducting phase transitions in thin layered structures.

\vfill\eject
\noindent
{\bf Acknowledgements}

The author M. N. wishes to acknowledge the hospitality of the Division of Theoretical Physics
at the Physics Department, University of Helsinki.


\begin{thebibliography}{9}
\bibitem{BCS} J. Bardeen, L. N. Cooper and J. R. Schrieffer, Phys. Rev. {\bf 108}, 1175 (1957)
\bibitem{Abrikosov} A. A. Abrikosov, L. P. Gor'kov and I. E. Dzyaloshinskii, Quantum Field
Theoretical Methods in Statistical Physics (Pergamon Press, London, 1963); A. L. Fetter and
J. D. Walecka, Quantum Theory of Many-Particle Systems (McGraw-Hill Inc., New York, 1971)
\bibitem{Meservey} R. Meservey and B. B. Schwartz, in Superconductivity, Vol. 1, p. 117, edited
by R. D. Parks (Marcel Dekker, Inc., New York, 1969)
\bibitem{Abrikosov 2} See e.g. A. A. Abrikosov, Fundamentals of the Theory of Metals 
(North-Holland, Amsterdam, 1988)
\bibitem{Landau} L. D. Landau and E. M. Lifshitz, Statistical Physics (Pergamon Press, Oxford,
1980) 
\bibitem{Gor'kov} L. P. Gor'kov, Zh. Eksp. Teor. Fiz. {\bf 37}, 833 (1959) [Sov. Phys. JETP 
{\bf 10}, 593 (1960)]
\bibitem{Bogolu} N. N. Bogoliubov, Physics {\bf 26} (1960)S 1.
\bibitem{Milan} P. Kalinay and M. Noga, Czech. J. Phys. {\bf 48} (1998), 1615
\bibitem{Burns} See e.g. G. Burns, High Temperature Superconductivity (Academic Ptress, Inc., 
Boston, 1992)

\end{thebibliography}
\end{document}